\documentclass[twocolumn,superscriptaddress,floatfix,amsmath,amssymb,aps]{revtex4-1}
\usepackage[utf8]{inputenc}
\usepackage{chemformula}
\usepackage{amsmath}
\usepackage{mathrsfs}
\usepackage{physics}
\usepackage{multirow}
\usepackage{amssymb}
\usepackage{graphicx}
\usepackage{siunitx}
\usepackage{lipsum}
\usepackage[normalem]{ulem}
\usepackage[colorlinks=true, allcolors=blue]{hyperref}
\usepackage{float}

\newcommand{\Tc}{$T_c$ }

\newcommand{\NbMoBdi}{Nb$_{1-x}$Mo$_x$B$_2$ }

\definecolor{mag}{RGB}{255,0,255}

\begin{document}
\title{High critical field superconductivity at ambient pressure in MoB$_2$ stabilized in the P6/mmm structure via Nb substitution}

\date{\today}

\begin{abstract}
Recently it was discovered that, under elevated pressures, MoB$_2$ exhibits superconductivity at a critical temperature, $T_c$, as high as \SI{32}{K}.
The superconductivity appears to develop following a pressure-induced structural transition from the ambient pressure R$\bar{3}$m structure to an MgB$_2$-like P6/mmm structure.
This suggests that remarkably high $T_c$ values among diborides are not restricted to MgB$_2$ as previously appeared to be the case, and that similarly high $T_c$ values may occur in other diborides if they can be coerced into the MgB$_2$ structure.
In this paper, we show that density functional theory calculations indicate that phonon free energy stabilizes the P6/mmm structure over the R$\bar{3}$m at high temperatures across the Nb$_{1-x}$Mo$_x$B$_2$ series. 
X-ray diffraction confirms that the synthesized Nb-substituted MoB$_2$ adopts the MgB$_2$ crystal structure. High magnetic field electrical resistivity measurements and specific heat measurements demonstrate that Nb$_{1-x}$Mo$_x$B$_2$ exhibits superconductivity with $T_c$ as high as \SI{8}{K} and critical fields approaching \SI{6}{T}.

\end{abstract}

\author{A. C. Hire}
\thanks{These authors contributed equally to this work.}
\email[Corresponding Author: ]{ajinkya.hire@ufl.edu}
\affiliation{Department of Materials Science and  Engineering, University of Florida, Gainesville, Florida 32611, USA}
\affiliation{Quantum Theory Project, University of Florida, Gainesville, Florida 32611, USA}
\author{S. Sinha}
\thanks{These authors contributed equally to this work.}
\email[Corresponding Author: ]{ajinkya.hire@ufl.edu}
\affiliation{Department of Physics, University of Florida, Gainesville, Florida 32611, USA}
\author{J. Lim}
\affiliation{Department of Physics, University of Florida, Gainesville, Florida 32611, USA}
\author{J. S. Kim}
\affiliation{Department of Physics, University of Florida, Gainesville, Florida 32611, USA}
\author{P. M. Dee}
\affiliation{Department of Materials Science and  Engineering, University of Florida, Gainesville, Florida 32611, USA}
\affiliation{Department of Physics, University of Florida, Gainesville, Florida 32611, USA}
\author{L.\ Fanfarillo}
\affiliation{Department of Physics, University of Florida, Gainesville, Florida 32611, USA}
\affiliation{Scuola Internazionale Superiore di Studi Avanzati (SISSA), Via Bonomea 265, 34136 Trieste, Italy}
\author{J. J. Hamlin}
\affiliation{Department of Physics, University of Florida, Gainesville, Florida 32611, USA}
\author{R. G. Hennig}
\affiliation{Department of Materials Science and  Engineering, University of Florida, Gainesville, Florida 32611, USA}
\affiliation{Quantum Theory Project, University of Florida, Gainesville, Florida 32611, USA}
\author{P. J. Hirschfeld}
\affiliation{Department of Physics, University of Florida, Gainesville, Florida 32611, USA}
\author{G. R. Stewart}
\affiliation{Department of Physics, University of Florida, Gainesville, Florida 32611, USA}

\maketitle

\section{Introduction}
The discovery of high-temperature superconductivity in the binary hydrides~\cite{Duan2014,Drozdov2015,Somayazulu2019} reignited the dream of achieving room-temperature superconductivity through conventional electron-phonon coupling.
Although these materials display \Tc values of upwards of \SI{100}{K}, very high pressures and complicated experimental setup are required to stabilize them.
Recent efforts in lowering the pressure values have resulted in theoretical~\cite{Sun2019,DiCataldo2021,Hilleke2022} and experimental \cite{Semenok2021} works where a third element stabilizes the clathrate-like cage of hydrogen atoms at much lower pressures.
Another promising path for achieving high-temperature superconductivity is to search for superconducting materials containing light elements.
The AlB$_2$-type diborides have been of particular interest since the discovery of (record-breaking at the time) ambient pressure conventional superconductivity (\Tc = \SI{39}{K}) in MgB$_2$~\cite{Nagamatsu2001}.

Recently, Pei \emph{et al.}~\cite{Pie_MoB2} discovered high pressure-induced superconductivity in MoB$_2$ (\Tc = \SI{32}{K} above P = \SI{60}{GPa}, where the structure changes from R$\bar{3}$m to P6/mmm).
The P6/mmm MoB$_2$ phase and the hydride superconductors, despite being  different chemical systems, share some similarities. Like the high-pressure superconducting hydrides\cite{Li_2016,Liu_2017}, P6/mmm MoB$_2$ does not occur on the ambient pressure DFT convex hull, and the perfect stoichiometric structure is dynamically unstable at low pressures.
Our own similar work on WB$_2$ (\Tc = \SI{17}{K} above P = \SI{50}{GPa})~\cite{Lim_WB2} has focused our interest in finding such superconductivity stabilized at room pressure in the metal diborides.
Since NbB$_2$ occurs with P6/mmm symmetry, prototype AlB$_2$ structure, space group 191 (in which MoB$_2$ under pressure  exhibits superconductivity at \SI{30}{K}), we investigated the phase stability of Nb$_{1-x}$Mo$_x$B$_2$ vs Mo concentration, by density functional theory (DFT) calculations and  experimentally, by arc-melting of the constituents and characterization measurements as discussed below.

An early experimental work by Kuz’ma~\cite{Kuzma1971}, in which samples were formed at 1400 \textdegree C, found that Nb$_{1-x}$Mo$_x$B$_2$ remained in the P6/mmm, AlB$_2$ structure beginning at pure NbB$_2$ ($x = 0$) and up to $x = 0.24$.
Kuz’ma went on to say that, based on their results, ``continuous series of solid solutions may form between the isostructural compounds NbB$_2$ - MoB$_2$'' at other temperatures.
According to the Materials Platform for Data Science, the melting point of NbB$_2$ is just over 3000 \textdegree C.
MoB$_2$ forms peritectally around 2350 \textdegree C~\cite{Storms1977}.
As will be seen below in the Results section, arc-melting (an intrinsically high temperature process) of the constituents in Nb$_{1-x}$Mo$_x$B$_2$ indicates that these pseudobinary compounds remain in the P6/mmm structure upon cooling all the way from pure NbB$_2$ up to  $x=0.9$.
Thus, the Al-flux grown MoB$_2$ of  Pei \emph{et al.}, which formed in the R$\bar{3}$m structure below 1500 \textdegree C and  transforms to P6/mmm above \SI{60}{GPa}, is readily stabilized at room pressure in the hexagonal P6/mmm AlB$_2$ structure with minor additions of Nb. 

This paper presents DFT calculations elucidating the effect of Nb addition and temperature on the stability of the P6/mmm MoB$_2$ phase.
Coupled with these calculations, we present x-ray diffraction, resistivity, specific heat at low temperatures, and upper critical magnetic field, H$_{\text{c2}}$(0) for arc-melted samples of Nb$_{1-x}$Mo$_x$B$_2$, x= 0.25, 0.50, 0.75, and 0.9---all of which show essentially single phase, P6/mmm x-ray patterns (with no indication of Nb second phase) and exhibit significant fractions of bulk superconductivity at ambient pressure.
Data for pure arc-melted stoichiometric NbB$_2$ (P6/mmm) and MoB$_2$ (R$\bar{3}$m) showed no superconductivity, in agreement with the paper by Fisk~\cite{Fisk1991}.

\section{Methods}
For performing the density functional theory calculations, we use VASP~\cite{PhysRevB.47.558,PhysRevB.49.14251,KRESSE199615,PhysRevB.54.11169}.
VASP calculations were performed using the projector augmented wave pseudopotentials \cite{PhysRevB.50.17953} and Perdew-Burke-Ernzerhof (PBE)~\cite{PhysRevLett.77.3865} generalized gradient approximation for the exchange-correlation functional.
A k-point density of 60 per \AA$^{-1}$ and plane wave cut-off of \SI{520}{eV} was used in all the calculations.
To estimate phonon entropy contributions to the free energy, we first calculate the phonon frequencies for a supercell containing 24 atoms in VASP using perturbation theory.
Then the phonon density of states was obtained from the phonon frequencies by using the Gaussian smearing method, similar to the one adopted by Lim,~\emph{et al.}~\cite{Lim2021}.
For smearing the electrons in the DFT calculations, we use the Methfessel-Paxton smearing with a $\sigma$ of \SI{0.1}{eV} for NbB$_2$, Nb$_{0.75}$Mo$_{0.25}$B$_2$, and Nb$_{0.5}$Mo$_{0.5}$B$_2$ P6/mmm phases and for R$\bar3$m MoB$_2$.
For P6/mmm Nb$_{0.25}$Mo$_{0.75}$B$_2$, and MoB$_2$ phases we use the Fermi smearing with smearing values of 0.1, 0.25, and \SI{0.5}{eV}, as discussed below.

For experimental measurements, \NbMoBdi ($x = 0.25, 0.5, 0.75, 0.9, 1$) samples were formed via arc melting the constituent elements. A reasonable estimate for the temperature range for arc melting the constituent elements is between 2400\textdegree C and 2700\textdegree C. Resistivity measurements were done on a bulk sample with dimensions $\sim$ $4.2 \times 3.5 \times 0.4$ mm$^3$, which was set up in Kelvin sensing configuration for 4-probe measurements. Quantum Design Physical Property Measurement System (PPMS) was used along with a Keithley 6221/2182A Delta Mode System for temperature control and voltage measurements. Current of \SI{1}{mA} was used for all the measurements on this sample. The sample was idealized to be rectangular with uniform thickness for the measurements. Small-scale errors from these assumptions were not taken into consideration.  Specific heat at low temperatures was measured using standard time constant methodology~\cite{Stewart1983}.

\section{Results and Discussion}
\subsection{DFT}
\begin{figure}
    \includegraphics[width=\linewidth]{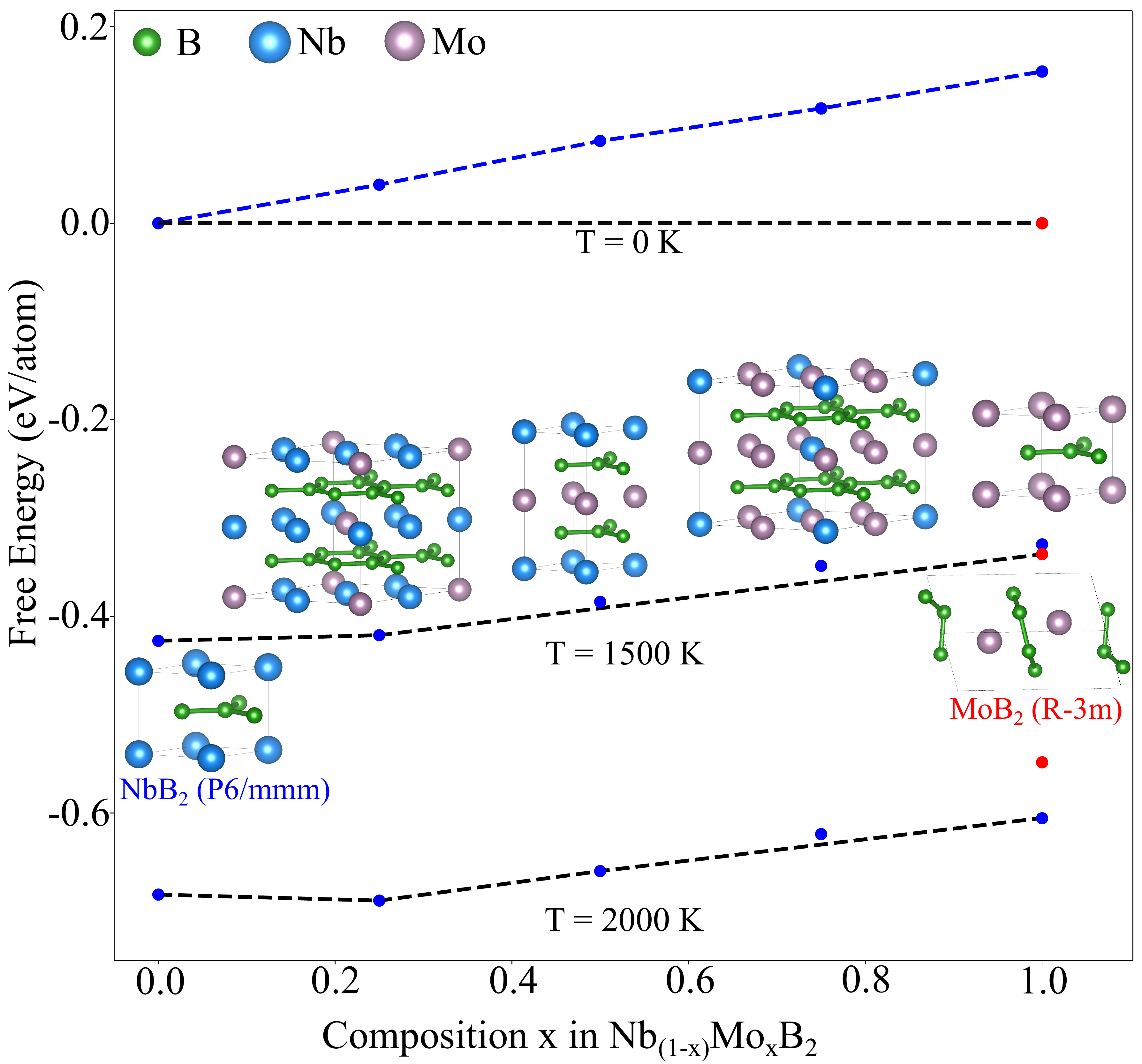}
    \caption{Calculated free energy of \NbMoBdi as a function of composition and temperature. The free energy was calculated with respect to 0K free energy of stable NbB$_2$ (P6/mmm) and MoB$_2$ (R$\bar3$m) phases. The blue dots represent the P6/mmm phase and the red dots represent the R$\bar3$m phase. Only vibrational entropy contribution to the free energy was considered. The black line is the convex hull line at various temperatures. Electron smearing of \SI{0.5}{eV} was used for the phonon calculation of P6/mmm MoB$_2$ and Nb$_{0.25}$Mo$_{0.75}$B$_2$.}
    \label{NbMoB2_form_eng}
\end{figure}

In this section, we will discuss the DFT calculated stability of the P6/mmm phase in \NbMoBdi alloys at \SI{0}{K} and high temperatures. At \SI{0}{K}, it is determined by the compound convex hull construction. To estimate stability at high temperatures, we calculate the free energy of the compound and its competing phases, considering only the phonon contributions to entropy as a function of composition and temperature. We calculate the phonon dispersion curves for P6/mmm \NbMoBdi ($x = 0.25, 0.5, 0.75, 1$) and for R$\bar3$m MoB$_2$ using the supercell method in VASP.
Figure~\ref{NbMoB2_form_eng} shows the DFT calculated free energy of \NbMoBdi as a function of composition at three temperatures, $T$ = \SI{0}{K}, \SI{1500}{K} and \SI{2000}{K}.
The figure also shows the structure of the supercells used in the DFT calculations. From this figure at \SI{0}{K}, the P6/mmm \NbMoBdi is not stable (for $x = 0.25, 0.5, 0.75, 1$) with respect to decomposition to the stable end-point phases---NbB$_2$ (P6/mmm) and MoB$_2$ (R$\bar 3$m).
\\
 In DFT calculations, electronic states near the Fermi energy are generally smeared out to help with the convergence of numerical integrals\cite{Sholl_2009,Lee_2016}. The smearing of electronic states is especially used for metals where the occupation of electrons discontinuously changes from 1 to 0 at the Fermi energy at $T =$ \SI{0}{K}. For low values of electron smearing ($\sigma =$ \SI{0.1}{eV}) typically used in the DFT calculations, stoichiometric P6/mmm MoB$_2$ and Nb$_{0.25}$Mo$_{0.75}$B$_2$ exhibit dynamical instabilities, meaning the calculated phonon dispersion curves have imaginary phonons. On the other hand the calculated phonons for Nb$_{0.75}$Mo$_{0.25}$B$_2$ and Nb$_{0.5}$Mo$_{0.5}$B$_2$ are real and the structures are dynamically stable, indicating that the P6/mmm phases can be stabilized at high temperatures by entropy. Stable phonons for the P6/mmm MoB$_2$ and Nb$_{0.25}$Mo$_{0.75}$B$_2$ phases can be obtained by using high smearing values ($\sigma =$ \SI{0.5}{eV}, fermi smearing) for the electrons in the DFT calculations. A similar scheme with smearing value as high as $\approx$\SI{2.7}{eV} was used by Babu and Guo~\cite{Babu2019} and Ivashchenko \emph{et al.}~\cite{Ivashchenko2010} for stabilizing the phonons of the delta phase NbN.

\begin{figure}
    \includegraphics[width=\linewidth]{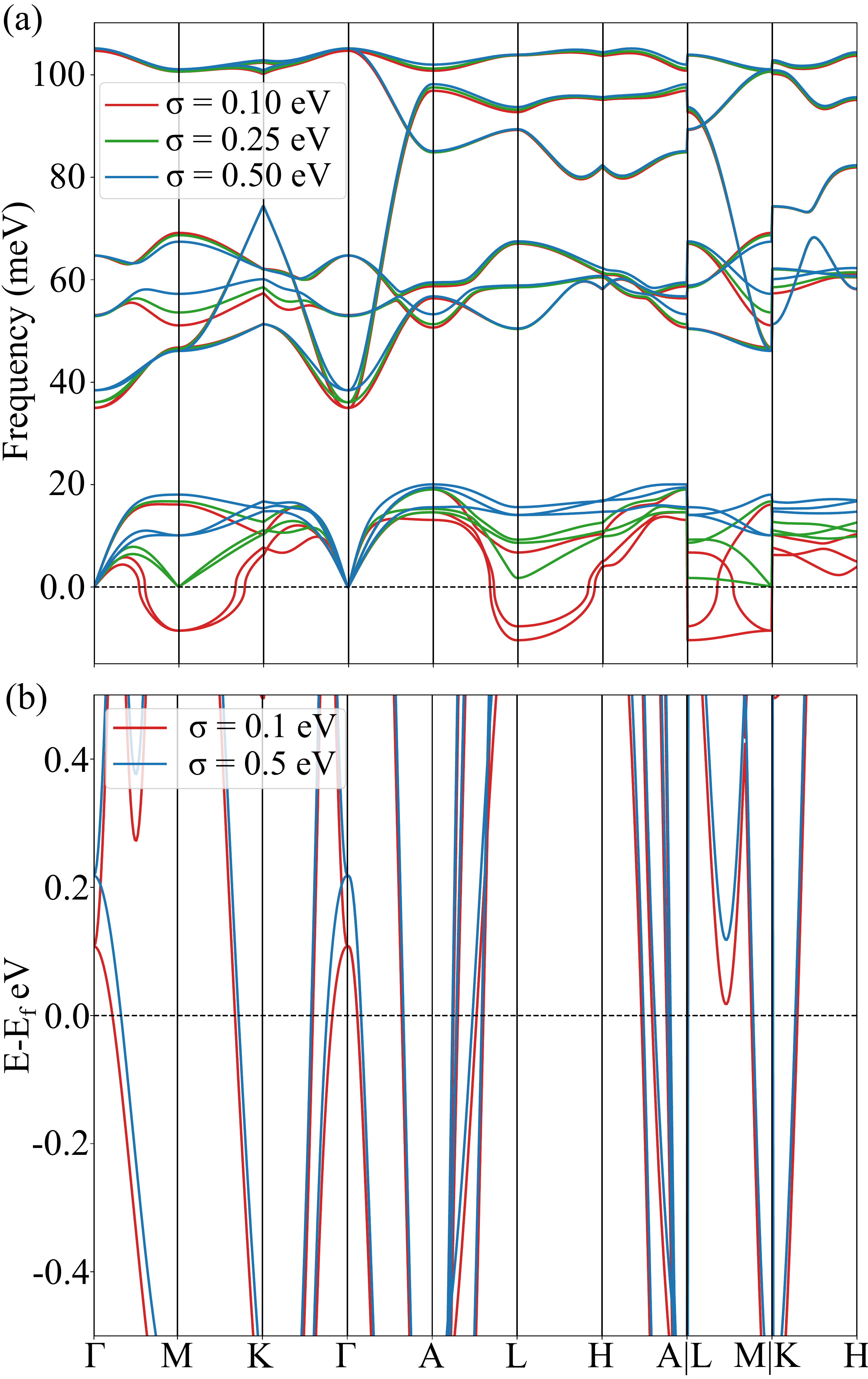}
    \caption{(a) P6/mmm MoB$_2$ phonon dispersion curves and (b) Bband structure diagram for varying electron smearing values. With increasing electron smearing the P6/mmm phase of MoB$_2$ becomes dynamically stable. Notice the downward shift of the Fermi level with increasing smearing.}
    \label{MoB2_ph_BS}
\end{figure}

Figure~\ref{MoB2_ph_BS} shows the phonon dispersion curves and band structure for MoB$_2$ P6/mmm phase at various electron smearing values. The Fermi smearing method was used in the DFT calculations. Imaginary acoustic phonon frequencies are present at the ``M'' and ``L'' high symmetry points and also along ``L-M'' path. From Fig.~\ref{MoB2_ph_BS}(a) the P6/mmm MoB$_2$ phase becomes dynamically stable at large electron smearing values, causing a downward shift of the Fermi energy (Fig.\ref{MoB2_ph_BS}(b)). A similar downward shift of Fermi energy can also be expected in structures with Mo vacancies. We observe an analogous trend of stable phonons for high smearing values in Nb$_{0.25}$Mo$_{0.75}$B$_2$. For $\delta$-NbN it has been argued that the large smearing values capture, to a first-order approximation, the effect of temperature and disorder due to the nitrogen vacancies present in the experimental delta phase NbN samples~\cite{Ivashchenko2010,Babu2019}.

Using the above-discussed smearing values for P6/mmm MoB$_2$ and Nb$_{0.25}$Mo$_{0.75}$B$_{2}$, we calculate the high-temperature free energy for P6/mmm \NbMoBdi alloys and R$\bar3$m MoB$_2$. Figure \ref{NbMoB2_form_eng} also shows the estimated compound convex hull for \SI{1500}{K} and \SI{2000}{K}. At higher temperatures, the doped P6/mmm \NbMoBdi ($x=0.25,0.5$) phases are on the convex hull. The  distance from the convex hull for Nb$_{0.25}$Mo$_{0.75}$B$_{2}$ also decreases and is close to stability at high temperatures. We believe that our Nb$_{0.25}$Mo$_{0.75}$B$_2$ arc-melted samples are dynamically stabilized by disorder. Thus, the P6/mmm \NbMoBdi is stabilized with respect to decomposition by the phonon entropy and can be obtained experimentally at high temperatures.

\subsection{X-ray diffraction}
\begin{figure}
    \includegraphics[width=\linewidth]{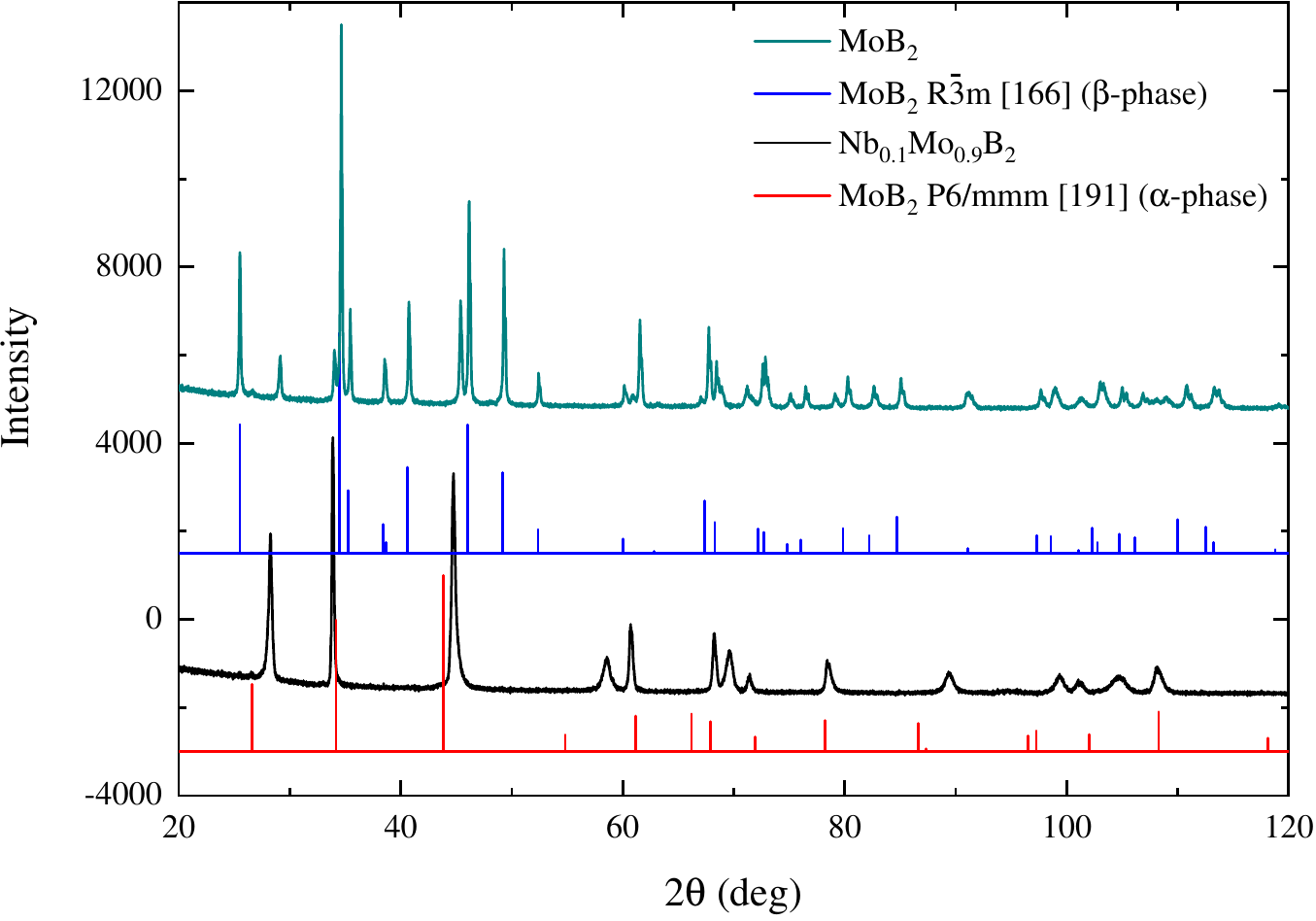}
    \caption{X-ray patterns of arc-melted MoB$_2$ (upper trace, green) and arc-melted Nb$_{0.1}$Mo$_{0.9}$B$_2$ (third trace from the top, black) compared to calculated patterns for R-3m structure MoB$_2$ (blue, space group 166, second trace) and for P6/mmm structure MoB$_2$ (red, space group 191, bottom trace.)  Although there are reflections in the measured patterns that do not agree with the calculated ones (e.g. there are extra lines at ~30.0$^{\circ}$ and ~45.4$^{\circ}$ 2$\theta$ in the arc-melted MoB$_2$ data which may be from a second phase of the P6/mmm structure), clearly the measured data support the conclusions that a.) arc-melted MoB$_2$ occurs primarily in the R-3m structure while $10\%$ Nb doping (Nb$_{0.1}$Mo$_{0.9}$B$_2$) causes the arc-melted sample to occur essentially entirely in the P6/mmm structure.}
    \label{exp_xrd_fig_1}
\end{figure}

\begin{figure}
    \includegraphics[width=\linewidth]{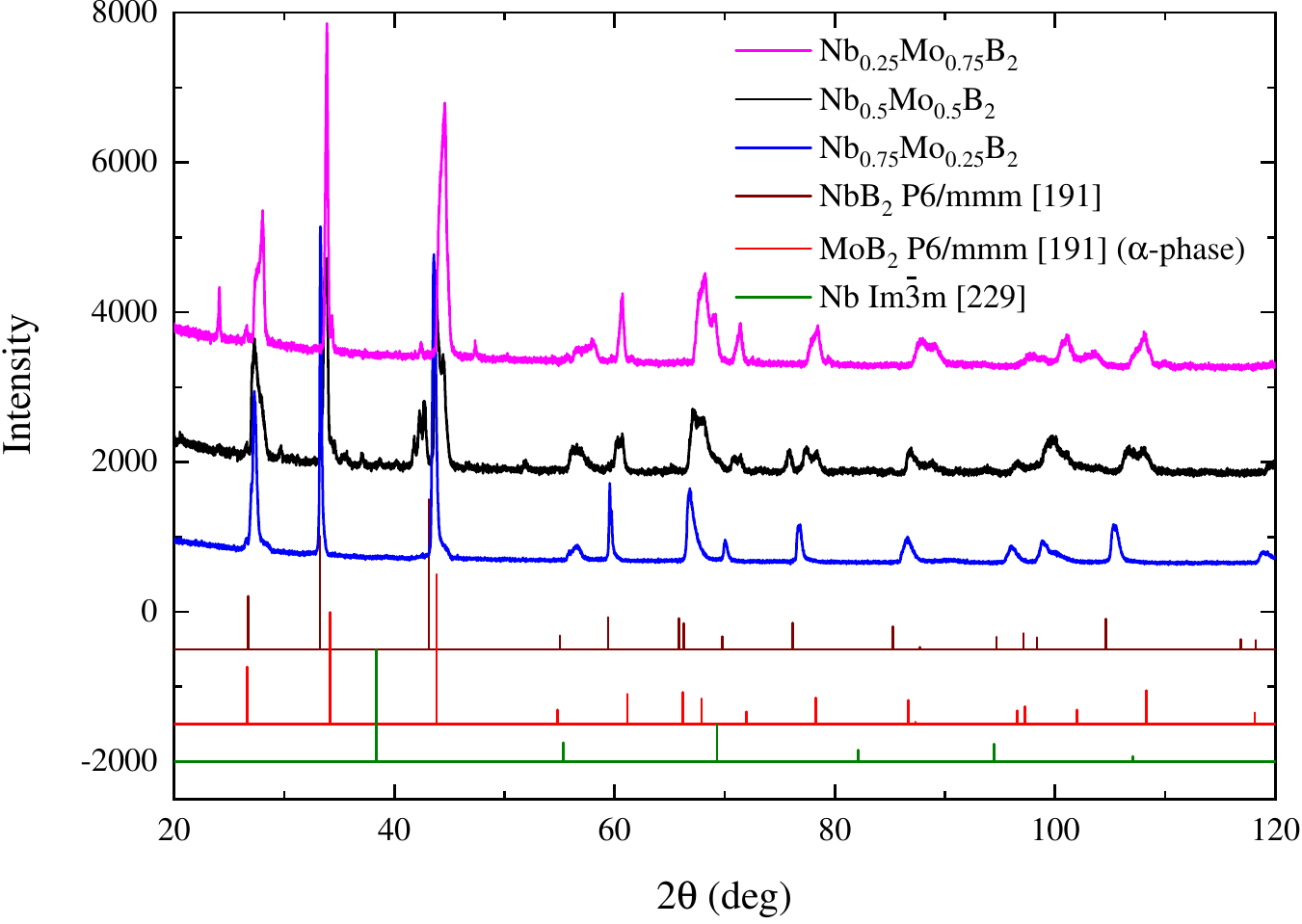}
    \caption{Measured x-ray diffraction patterns for arc-melted Nb$_{1-x}$Mo$_x$B$_2$, $x=$ 0.25, 0.50, and 0.75, and calculated patterns for both MoB$_2$ and NbB$_2$ in the P6/mmm structure, as well as for pure Nb.  Again, all of the \NbMoBdi samples, $x=$ 0.25, 0.50, and 0.75, occur in the P6/mmm structure, with no evidence of second phase Nb.  Two second phase lines are, however, observed.   In Nb$_{0.25}$Mo$_{0.75}$B$_2$ there is a second phase line at 23.9$^{\circ}$ 2$\theta$; in Nb$_{0.5}$Mo$_{0.5}$B$_2$ there is a second phase line at 42.65$^{\circ}$ 2$\theta$. As shown, neither of these lines match the Nb x-ray pattern.}
    \label{exp_xrd_fig_2}
\end{figure}

X-ray diffraction (XRD) results for arc-melted MoB$_2$ and \NbMoBdi, $x = 0.9$, along with the calculated x-ray patterns for the low temperature structure R$\bar3$m of MoB$_2$ and the high temperature P6/mmm structure for MoB$_2$, are shown in Fig.~\ref{exp_xrd_fig_1}.  Clearly, the arc-melted MoB$_2$ occurs in the R$\bar3$m structure, while only $10\%$ Nb stabilizes Nb$_{0.1}$Mo$_{0.9}$B$_2$ in the P6/mmm structure. Figure \ref{exp_xrd_fig_2} shows data for Nb$_{1-x}$Mo$_x$B$_2$, $x= 0.25, 0.50,$ and $0.75$, and pure Nb, and calculated patterns for both MoB$_2$ and NbB$_2$ in the P6/mmm structure.  The \NbMoBdi samples, $x= 0.25, 0.50,$ and $0.75$, occur in the P6/mmm structure, with no evidence of second phase Nb.

The lattice parameters for all four \NbMoBdi, $x= 0.25, 0.50, 0.75,$ and $0.9$ as well as for the calculated end compositions, NbB$_2$ and MoB$_2$, in the hexagonal P6/mmm structure (as well as measured parameters discussed below) are shown in Table \ref{table_1}. As can be seen, the substitution of Mo by Nb does not lead to large lattice parameter changes.

\begin{table*}
\begin{tabular}{c|c|c|c|c|c|c|c|c|c|c|c|c|c}
\hline \hline
\multirow{2}{*}{Material} & \multicolumn{2}{c|}{From $\rho$ vs $T$ (K)} & \multirow{2}{*}{$\frac{\Delta C}{\gamma T_c}$}& \multirow{2}{*}{RRR} & \multicolumn{2}{c|}{Calculated \AA} & \multicolumn{2}{c|}{Measured \AA} & $H_{c2}$ & $\gamma$ & DOS & $\beta$ & $\Theta_D$ \\
                          & $T_c^{\text{onset}}$ & $T_c^{\text{mid}}$ &  &   & $a$ & $c$ & $a$ & $c$ & (T) & (mJ/molK$^2$) &  & (mJ/molK$^4$) & (K)\\
\hline \hline
MoB$_2$                     &       &       &       & 2.74 & 3.028  & 3.3239 &  &  &                &       & 1.5213      &        &     \\
Nb$_{0.1}$Mo$_{0.9}$B$_2$   & 7.82  & 6.91  & 0.32  & 1.11 &        &        & 3.0579 & 3.3533 & 6.3$\pm$0.3    &  3.31 &           & 0.0124 & 715  \\
Nb$_{0.25}$Mo$_{0.75}$B$_2$ & 8.15  & 6.85  & 0.99  & 1.08 & 3.051 & 3.3323 & 3.0546 & 3.2637 & 6.7$\pm$0.3    &  3.80 & 1.36      & 0.0142 & 625  \\
Nb$_{0.5}$Mo$_{0.5}$B$_2$   & 6.51  & 5.25  & 0.44  & 1.15 & 3.0918 & 3.291 & 3.0812 & 3.2989 & 5.1$\pm$0.3    &  3.27 & 1.26      & 0.0128 & 693  \\
Nb$_{0.75}$Mo$_{0.25}$B$_2$ & 6.98  & 5.88  & 0.27  & 1.31 & 3.100 & 3.3243 & 3.1053 & 3.3026 & 2.2$\pm$0.3    &  2.88 & 0.88      & 0.0105 & 845  \\
NbB$_2$                     &       &       &       &      & 3.1186 & 3.3385 & 3.1024\cite{Mudgel_2008} & 3.3196\cite{Mudgel_2008} &                &  1.99 & 1.06      & 0.0096 & 846  \\
\hline
\end{tabular}
\caption{Parameters for \NbMoBdi P6/mmm phase, $x= 0, 0.25, 0.50, 0.75, 0.9$ and $1$. DOS is states per eV per formula unit.}
\label{table_1}
\end{table*}

\subsection{Resistivity}
The onset \Tc in resistivity measurement for Nb$_{0.25}$Mo$_{0.75}$B$_2$ was found to be $\sim$\SI{8}{K} in accordance with specific heat measurements. All the other samples (\NbMoBdi, $x= 0.25, 0.50,$ and $0.9$) show a onset \Tc in resistivity measurement of less than $\SI{8}{K}$. 
The residual resistivity ratios [$R$(300 K)/$R(T _{c+}$)] for $x= 0.25, 0.5, 0.75,$ and $0.9$ are all between 1.08 and 1.31, reflecting the scattering caused by alloying with Nb. Pure MoB$_2$ has RRR = 2.74.
 
The $T_c$’s obtained for each Nb$_{x}$Mo$_{1-x}$B$_{2}$ compound are around four times smaller than the \SI{32}{K} observed in MoB$_2$ under high pressure (\SI{109.7}{GPa}) \cite{Pie_MoB2} and on par with $T_{c}\sim$\SI{9}{K} in nonstoichiometric NbB$_2$ \cite{Yamamoto2002,Mudgel_2008}. Stoichiometric NbB$_2$, by recent accounts, is not a superconductor \cite{Gasparov_2001,Mudgel_2008,Mudgel_2009}. The finite $T_c$ observed in the non-stoichiometric compounds containing either excess boron or niobium vacancies appears alongside a notable expansion of the $c$-axis lattice parameter. Shifts in the lattice parameter can indicate changes in the electronic structure, lattice vibrations, and electron-phonon coupling that ultimately affect superconductivity. Among these potential changes, increasing the electron-phonon coupling to boron-$p$ optical phonons offers a disproportionately large contribution to $T_c$ in MoB$_{2}$~\cite{Quan2021}. In Nb$_{x}$Mo$_{1-x}$B$_{2}$, however, Nb primarily helps stabilize the P6/mmm structure at low pressures and does not recreate other conditions needed for an even higher $T_c$ observed in MoB$_2$ under high pressure.

\subsection{Specific Heat}
\begin{figure}
    \includegraphics[width=\linewidth]{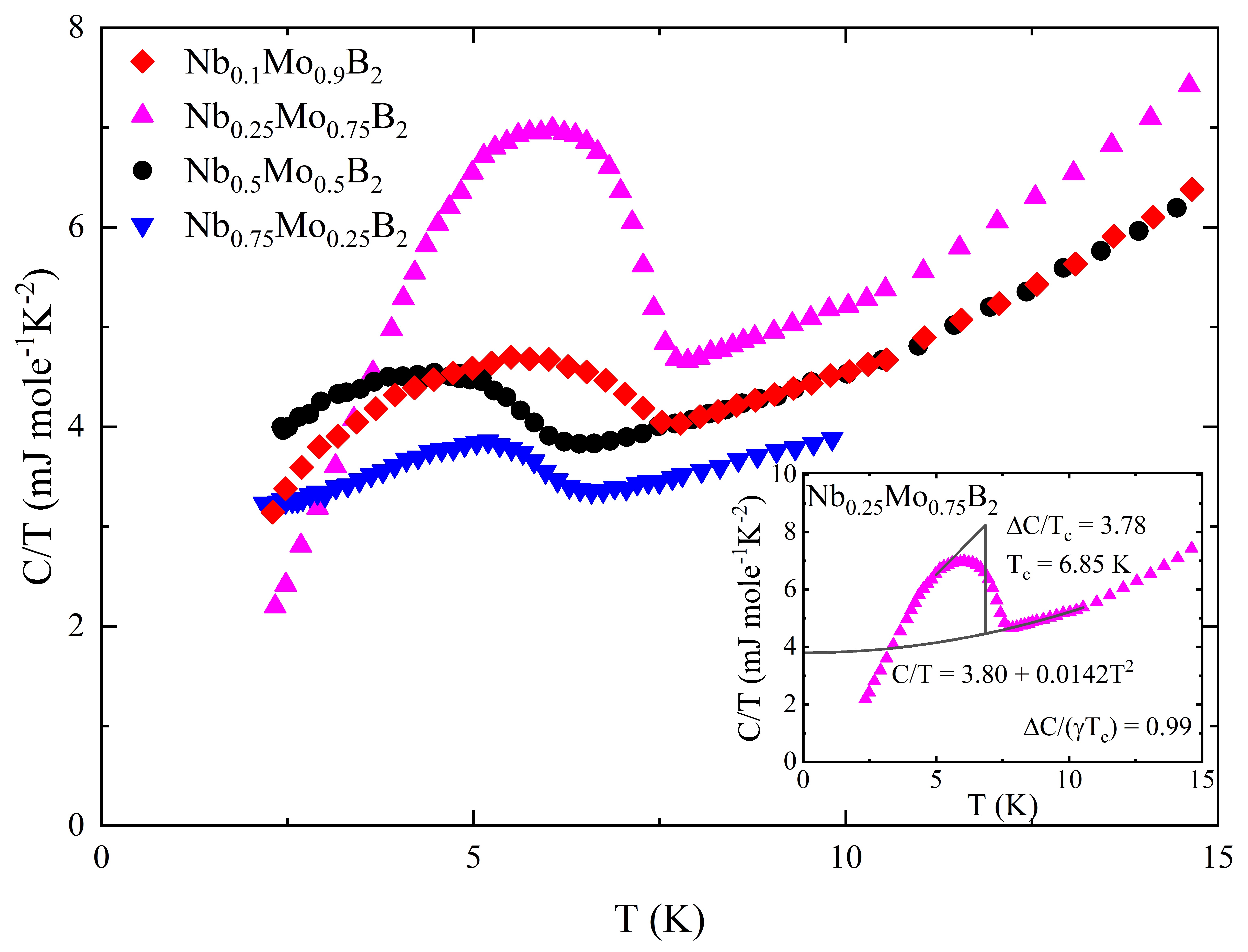}
    \caption{Specific heat data of \NbMoBdi, $x= 0.25, 0.50, 0.75, 0.9$.}
    \label{exp_specific_heat}
\end{figure}
The specific heat data of \NbMoBdi, $x= 0.25, 0.50, 0.75, 0.9$ are shown in Fig.~\ref{exp_specific_heat}. The derived values for $\gamma$ [$=\lim_{T\to 0} C_{\text{normal}}/T$, where $\gamma \propto$ N(0), the electronic density of states at the Fermi energy] and $\Delta C/(\gamma T_c)$ (a measure of the amount of bulk superconductivity, with $\Delta C/(\gamma T_c)=1.43$ for BCS superconductivity or 1.64 for the unconventional, iron-based superconductor\cite{Kim_2015} FeSe \Tc=\SI{8.1}{K}) are listed in Table \ref{table_1}. Based on the rough approximation that $\Delta C/(\gamma T_c)\approx1.5$ for $100\%$ bulk superconductivity, the specific heat data show that alloying MoB$_2$ with Nb and stabilizing the P6/mmm structure at ambient pressure  creates between $20\%$ to $70\%$ bulk superconductivity in the pseudobinary alloys, with $T_c$’s between \SI{6.5}{K} and \SI{8.15}{K}. Since R$\bar3$m structure MoB$_2$ was found \cite{Pie_MoB2} to be superconducting above \SI{60}{GPa} after transforming to P6/mmm, it seems likely that the small amounts of Nb in our sample simply act to stabilize the favorable structure for superconductivity.  We note that these samples show no evidence of any second phase of Nb in the x-ray diffraction patterns (Figs.~\ref{exp_xrd_fig_1} and \ref{exp_xrd_fig_2}). Thus, these results indicate that Nb alloying into MoB$_2$ has succeeded in stabilizing superconductivity at ambient pressure in the same structure (P6/mmm) where Pei \emph{et al.} found~\cite{Pie_MoB2} superconductivity at high pressure, with the largest fraction of induced superconductivity occurring at only $25\%$ Nb content. We suspect that the higher $T_c$ value shown by our Nb$_{0.25}$Mo$_{0.75}$B$_2$ sample might be because of certain aspects of the ordering of the Nb and Mo atoms in the sample.

The chemical homogeneity of our samples can be inferred from the width of high angle XRD lines (Fig. \ref{exp_xrd_fig_2}) and from the width of the specific heat anomaly at the the transition (Fig. \ref{exp_specific_heat}). The width of the specific heat anomaly $\Delta C$ clearly indicates a distribution of concentration for all our samples. The highest \Tc alloy, Nb$_{0.25}$Mo$_{0.75}$B$_2$, perhaps demonstrates a lower level of homogeneity as indicated by the wide high-angle XRD lines (at $\approx$108\textdegree, full width half maximum $\approx$1\textdegree) and specific heat anomaly, as compared to the homogeneity of Nb$_{0.5}$Mo$_{0.5}$B$_2$, and Nb$_{0.75}$Mo$_{0.25}$B$_2$.

\subsection{Upper Critical Field}
\begin{figure}
    \includegraphics[width=\linewidth]{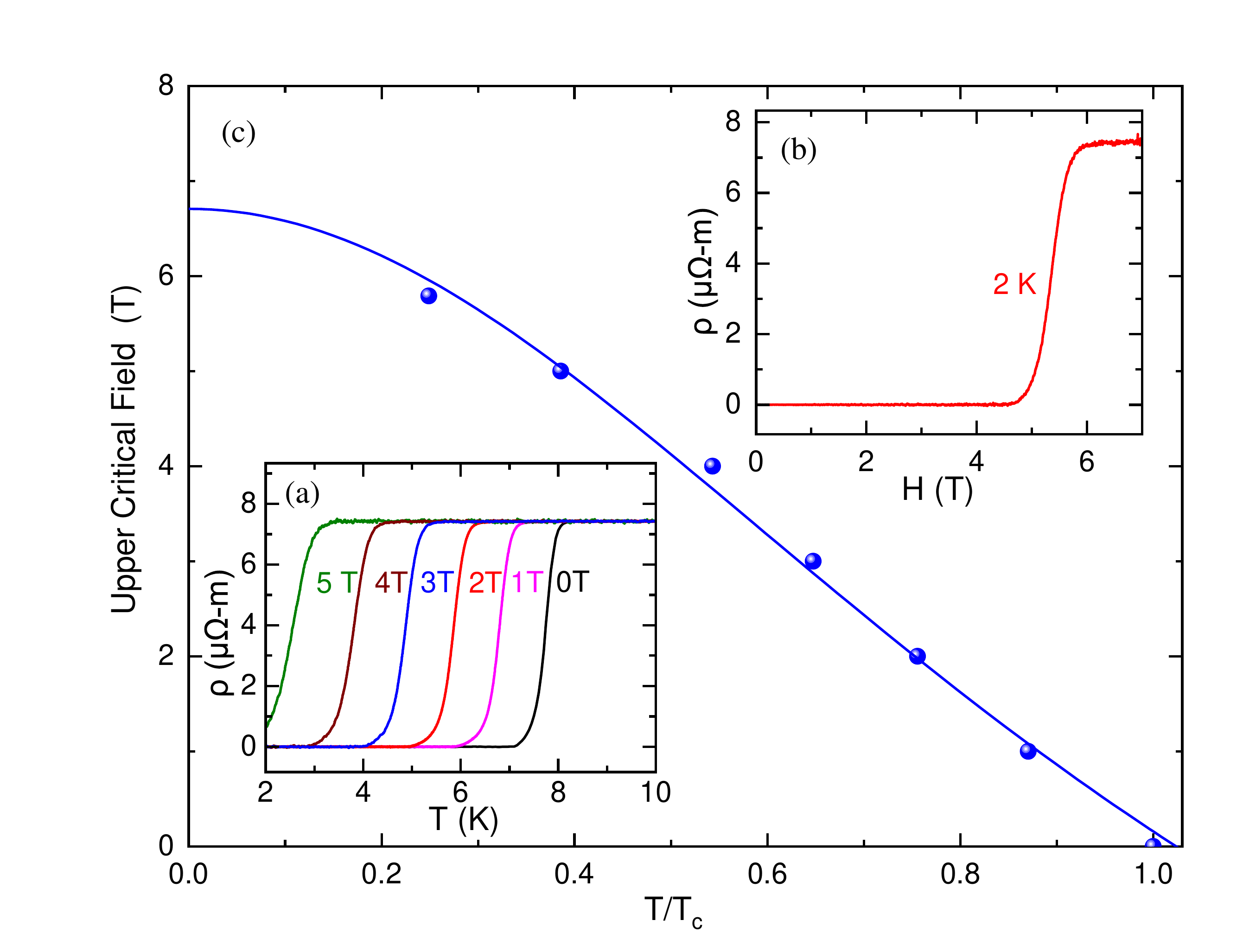}
    \caption{Resistivity dependence of Nb$_{0.25}$Mo$_{0.75}$B$_2$ (a) \textcolor{mag}{on} temperature for different magnetic fields, and (b) on magnetic field at constant temperature of \SI{2}{K}. T$_c$ is found to be \SI{8.03}{K}. (c) Upper critical field values for H$_{c2}$(0) for Nb$_{0.25}$Mo$_{0.75}$B$_2$ extracted from (a) and (b). The solid line shows the fitting to Ginzburg-Landau formula given by: $H_{c2} = H_{c2}(0)[(1-t^2)/(1+t^2)]$, where $t$ is the temperature normalized by the zero $-$ field transition temperature~\cite{Pie_MoB2}. This fitting gives $H_{c2}$(0) of \SI{6.71}{T}.}
    \label{fig_6}
\end{figure}
Figure~\ref{fig_6}(a) shows the suppression of transition temperature with applied field in Nb$_{0.25}$Mo$_{0.75}$B$_2$.
Resistivity dependence of Nb$_{0.25}$Mo$_{0.75}$B$_2$ on magnetic field at 2 K is shown in Fig.~\ref{fig_6}(b).
In Fig.~\ref{fig_6}(c), the experimental data are fit to the empirical formula given by $H_{c2} = H_{c2}(0)[(1-t^2)/(1+t^2)]$, where, $t = T/T_{c}$~\cite{Pie_MoB2}. 
Using this fit, the upper critical field at T = 0, H$_{c2}(0)$, is found to be \SI{6.71}{T}.
This unusually large critical field of Nb$_{0.25}$Mo$_{0.75}$B$_2$ is interesting. There are other conventional superconductors with a relatively large upper critical field to T$_c$ ratio. For example NbC has a $H_{c2}\approx$\SI{2}{T}\cite{Shang_2020} (T$_c =$ \SI{11.5}{K}) and NbTi has a H$_{c2}\approx$\SI{15}{T}\cite{Godeke_2007} (T$_c =$ \SI{9}{K}).
Pei \emph{et al.}, in their measurements of pure MoB$_2$, transformed by high pressure to form in the P6/mmm structure with a T$_c$ of about \SI{32}{K}, found H$_{c2}$(0) of only \SI{9.4}{T} whereas the \SI{32}{K} T$_c$ is a factor of four higher than that in Nb$_{0.25}$Mo$_{0.75}$B$_2$.
Our paper~\cite{Lim_WB2} on WB$_2$ under high pressure which, due to defects, has a similar structure to the P6/mmm, showed H$_{c2}$(0) = \SI{2.4}{T} and \Tc = 17 K, again arguing for Nb$_{0.25}$Mo$_{0.75}$B$_2$ --- due to its unusually high H$_{c2}$(0).

\section{Conclusions}
In summary, we were able to successfully stabilize the high pressure P6/mmm phase of MoB$_2$ at ambient pressure by doping with Nb. The Nb-stabilized MoB$_2$ P6/mmm samples are superconducting with the highest T$_c$ of \SI{8}{K} for Nb$_{0.25}$Mo$_{0.75}$B$_2$. Density functional theory calculations confirm that the favorable P6/mmm structure, also believed responsible for \SI{17}{K} superconductivity in WB$_2$ \cite{Lim_WB2}, is likely to be stabilized entropically at ambient pressure.  The doping of other borides by Nb may therefore also help to stabilize the P6/mmm  and lead to higher transition temperatures.

We note in closing that the relatively high H$_{c2}(0)\simeq$\SI{6.7}{T} of Nb$_{0.25}$Mo$_{0.75}$B$_2$ (Nb has a similar T$_c$ with an $H_{c2}$ of only \SI{0.44}{T}) makes it a suitable candidate for coating the inside of Nb superconducting radio-frequency particle accelerators. This high H$_{c2}$ coating could help further improve cavity performance.  

\section*{Acknowledgments}
Work at the University of Florida was performed under the auspices of U.S. Department of Energy Basic Energy Sciences under Contract No.\ DE-SC0020385. and under the auspices of the U.S. National Science Foundation, Division of Materials Research under Contract No. \ NSF-DMR-2118718.  A.C.H.\ and R.G.H.\ acknowledge support from the Center for Bright Beams, U.S. National Science Foundation Award No. PHY-1549132.

\newpage
\bibliography{ref}

\end{document}